\documentclass[preprint]{vgtc}               

\ifpdf
  \pdfoutput=1\relax                   
  \usepackage{graphicx}                
  \DeclareGraphicsExtensions{.pdf,.png,.jpg,.jpeg} 
\else
  \usepackage{graphicx}                
  \DeclareGraphicsExtensions{.eps}     
\fi%

\graphicspath{{figures/}{pictures/}{images/}{./}} 

\usepackage{microtype}                 
\PassOptionsToPackage{warn}{textcomp}  
\usepackage{textcomp}                  
\usepackage{mathptmx}                  
\usepackage{times}                     
\usepackage{cite}                      
\usepackage{tabu}                      
\usepackage{booktabs}                  
\usepackage{todonotes}
\usepackage{enumitem}
\onlineid{4009}

\vgtccategory{Research}

\vgtcinsertpkg

\usepackage{subcaption}

\ieeedoi{xx.xxxx/TVCG.201x.xxxxxxx}

\title{Evaluating Situated Visualization in AR with Eye Tracking}

\author{Kuno Kurzhals\thanks{e-mail: Kuno.Kurzhals@visus.uni-stuttgart.de}\\ %
        \scriptsize University of Stuttgart %
\and Michael Becher\thanks{e-mail: Michael.Becher@visus.uni-stuttgart.de}\\ %
     \scriptsize University of Stuttgart %
\and Nelusa Pathmanathan\thanks{e-mail: Nelusa.Pathmanathan@visus.uni-stuttgart.de}\\ %
    \scriptsize University of Stuttgart %
\and Guido Reina\thanks{e-mail: Guido.Reina@visus.uni-stuttgart.de}\\ %
    \scriptsize University of Stuttgart %
     }

\teaser{
  \centering
  \includegraphics[width=\linewidth]{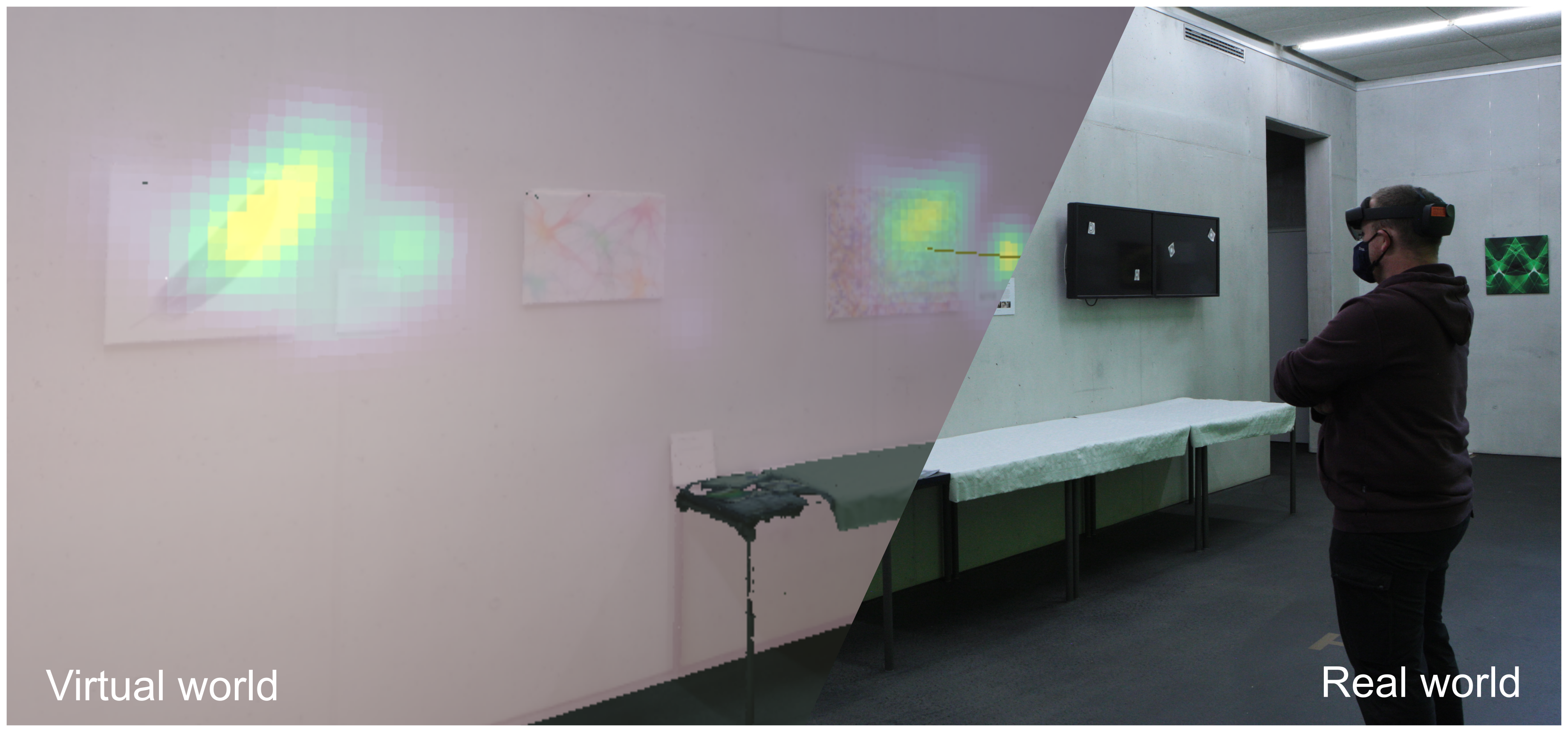}
  \caption{Example of recorded eye tracking data with an augmented reality device. Participants can interact with their environment in the real world while position and gaze are captured. The data is later represented with heat maps in a virtual model of the scene.}
  \label{fig:teaser}
}

\abstract{
Augmented reality (AR) technology provides means for embedding visualization in a real-world context. Such techniques allow situated analyses of live data in their spatial domain. However, as existing techniques have to be adapted for this context and new approaches will be developed, the evaluation thereof poses new challenges for researchers. Apart from established performance measures, eye tracking has proven to be a valuable means to assess visualizations qualitatively and quantitatively. We discuss the challenges and opportunities of eye tracking for the evaluation of situated visualizations. We envision that an extension of gaze-based evaluation methodology into this field will provide new insights on how people perceive and interact with visualizations in augmented reality. 
} 

\CCScatlist{
  \CCScatTwelve{Human-centered computing}{Visu\-al\-iza\-tion}{Visualization design and evaluation methods}{}
}

\begin{document}


\firstsection{Introduction}

\maketitle

Technical developments in recent years now provide means to bring visualization applications from the desktop out into the real world. 
Typical examples comprise techniques on mobile devices which augment data visualizations on real objects related to the data domain (e.g., urban planning~\cite{White2009}).
With the latest generation of head-mounted displays (HMDs), gesture-based interaction and hands-free movement become feasible and augmented content is provided directly in the user's field of view~\cite{Billinghurst2015, Hertel2021}.
Furthermore, an increasing number of devices is equipped with eye tracking technology. This allows measuring where the user's gaze is directed to, mainly as an intuitive input modality for human-computer interaction.
Examples include applications to inform about visual attention in collaborative scenarios~\cite{Billinghurst2002,Gupta2016,Jansen2020} and human-robot interaction~\cite{Chadalavada2020,Green2008}.
Similar applications are also possible in the context of situated data analysis~\cite{Sereno2020}, providing new means to interact with data in a natural way, by looking at it.
This information can be used directly for pointing, or as a measurement over time to predict behavior.

We argue that eye tracking is also beneficial for evaluation purposes in this context, especially for insights into spatial distributions of visual attention and problem-solving strategies. 
An example is displayed in Figure~\ref{fig:teaser}. The data was recorded to showcase the current state of the art for displaying gaze data in 3D environments. A person walks through a gallery and investigates different artworks. By recording the person's position and gaze during this procedure, post-experimental analysis can be performed in numerous ways. In the presented example, heat maps display distributed gaze measurements on a virtual model of the room.
The heat map shows where the person was looking, highlighting areas that attracted much visual attention. Such a simple visualization can already help evaluate the scenario, for instance, by identifying regions of importance that received little or no attention at all.

In the past, gaze-based analysis of participants working with visualizations has proven to be a strong supplement to established evaluation metrics~\cite{Kurzhals2014, Kurzhals2016}. Eye tracking provides qualitative and quantitative measures to inform about visual attention and cognitive processes~\cite{Holmqvist2011}.
To this point, the main application of eye tracking for the evaluation of visualizations lies in desktop-based scenarios. It is applied mainly in controlled lab studies, where typically one person sitting in front of a computer is tracked with a static device (i.e., a remote eye tracker).
Eye tracking glasses provide more flexibility but are pure recording devices and increase the analysis effort due to individual video recordings that often require extensive annotation work to derive meaningful insights.
With eye tracking integrated into HMDs, it is now possible to combine both aspects: 
\begin{itemize}
    \item Provide virtual content (e.g., data visualizations) and measure gaze distributions on it. This can be achieved with 2D and 3D visualizations, depending on the respective content.
    \item Embed the content in real-world environments where people are mobile and visualizations are presented in the spatial context of the data domain, a typical requirement for situated visualizations~\cite{Thomas2018}.
\end{itemize}

In this work, we discuss the challenges of eye-tracking based evaluation of situated visualizations and outline possible scenarios for future research directions. With the increasing availability of HMDs with eye-tracking capabilities, we see great potential for extending existing evaluation methodologies in the context of augmented reality and situated visualization.

\section{Situated Visualization}
Techniques for situated visualizations have been presented for the last decades on mobile devices and HMDs.
Especially the current generation of frameworks and development environments provide significantly increased convenience for designing visualizations embedded in real-world contexts~\cite{Fleck2022}.
One issue with the assessment of these new techniques is that especially quantitative evaluation procedures can rarely be translated from studies on desktop PCs into AR directly.

\subsection{What is situated visualization (SV)?}
Situated visualization describes the concept of bringing visual data analysis to the physical locations that the data refers to.
In doing so, the analysis process is no longer physically---and often not even temporally---removed from the investigated setting and it becomes possible to explore data in its natural spatial context.
This entails the use of portable, self-sufficient computing hardware such as smartphones, tablet computers, or augmented reality HMDs, on which the data is presented.
Thus, the growing interest in SV is closely linked to the progress made in ubiquitous computing and augmented reality in recent years.
Situated visualization is also closely related to the field of immersive analytics~\cite{Fonnet2021}, which focuses on exploring data in immersive environments, but without necessarily having these environments coincide with physical locations related to the data.

The use of situated visualizations offers various potential benefits:
It can, for example, provide bespoke data analysis to support on-site users in complex environments.
Without SV, they rely on knowledge gained during off-site data analysis sessions from which only high-level findings are easily remembered.
If data is available or brought on-site, access is often limited to traditional fixed displays or even analog media. Both of which lack direct spatial mapping to data referents and larger quantities of data cannot be properly processed without more advanced analysis tools.
Especially in live monitoring scenarios, an SV setup is beneficial to observe both the points of interest in the physical world and the data visualizations at the same time~\cite{Becher2022}.
Situated visualization can support engagement and also improve analysis in general by directly presenting data in the spatial context of the real world. Otherwise, this connection has to be made mentally, which increases cognitive load and potentially slows down the analysis process.

While the majority of current research on situated visualization focuses on augmented reality for delivery, alternatives such as projection mapping and data physicalization are also used.
For a wider coverage of SV methods and additional details, we refer to the survey on the current state of situated visualization by Bressa~et~al.~\cite{Bressa2022}.

\subsection{How is SV typically evaluated?}

The evaluation of AR systems proves to be difficult due to the variety of existing interfaces.
To gain meaningful and reliable results, it is important to select evaluation methods depending on the research question~\cite{Greenberg2008}. 
However, there is a lack of universal guidelines describing the design and conduction of an evaluation that researchers can follow in this case. 

In a systematic literature review, Merino et al.~\cite{Merino2020} classified different 
AR applications into seven different evaluation scenarios, derived from the work of Lam et al.~\cite{Lam2012} and Isenberg et al.~\cite{Isenberg2013}:
\begin{enumerate}[noitemsep]
    \item Algorithm performance
    \item Qualitative results inspection
    \item User performance
    \item User experience
    \item Understanding environment and work practices
    \item Team communication in AR
    \item Team collaboration in AR
\end{enumerate}

The scenarios mentioned above were further categorized into technique-centered, user-centered, and team-centered scenarios. Technique-centered evaluation methods exclude users and compare the performance and quality of novel techniques against existing benchmark algorithms, or include experts to comment and evaluate the system \cite{Duenser2011}.
User-centered scenarios are interested in the performance and experience of users interacting with the system. Evaluations measure the completion time and the correctness of user tasks, and also collect subjective feedback through Likert scale questionnaires and interviews.
Team-centered evaluation is based on assessing how well collaboration and communication works in the system. A systematic literature review that classified papers into the mentioned scenarios is provided by Merino and colleagues~\cite{Merino2020}. 
We argue that eye tracking can support all user-centered scenarios to derive additional measures for evaluation purposes, especially to gain insights into the cognitively complex scenarios 5--7.

\subsection{How would SV benefit from eye tracking?}
Behavior analysis benefits from the use of eye tracking for years~\cite{Hayhoe2005}. The recorded data provides information about how participants solve tasks from an egocentric view by indicating their viewing patterns. 
The inclusion of eye tracking measurements for situated visualization could provide deeper insights into cognitive processes and common strategies for interaction with visualizations, similar to desktop scenarios, i.e., estimating
the distribution of visual attention and gaze sequences for visual strategy analysis~\cite{Kurzhals2016}.
With the integration of eye tracking in the HMDs, this measure basically comes for free during the ex\-pe\-ri\-ment. Support software for gaze recording~\cite{Kapp2021} further facilitates this procedure. Hence, data acquisition is less problematic and the main question in this context is: 
\begin{center}
    ``\textit{How to analyze this bulk of spatio-temporal data?}''
\end{center}

\begin{table*}[ht]
\centering
\caption{Summary of benefits, challenges, and limitations for eye tracking in AR.}
\label{tab:challenges_summary}
\resizebox{\textwidth}{!}{%
\begin{tabular}{m{1.9cm}|m{4.5cm}|m{7.3cm}|m{6cm}}
 &
 
  \textbf{Benefits} &
  \textbf{Challenges} &
  \textbf{Limitations} \\ \hline
\textbf{Semantic Mapping} &
  \begin{itemize}[noitemsep,leftmargin=*]
      \item Context-aware visualizations
      \item Automatic labeling
  \end{itemize} &
  \begin{itemize}[noitemsep,leftmargin=*]
      \item Predefined virtual AOIs necessary for automatic labeling  
      \item Classification of gaze on real-world AOIs
  \end{itemize} &
  If world knowledge is limited, high effort to annotate real-world AOIs \\ \hline
\begin{tabular}[c]{@{}l@{}}\textbf{Spatial}\\\textbf{Mapping}\end{tabular}&
  \begin{itemize}[noitemsep,leftmargin=*]
      \item Show visualizations in spatial context of the real environment
      \item Comparability of tracking data recorded in the same environment
  \end{itemize} &
   \begin{itemize}[noitemsep,leftmargin=*]
  \item Simultaneous Localization and   Mapping (SLAM) for multiple recordings 
  \item Include virtual model (e.g., laser scan, photogrammetry, CAD models) 
  \end{itemize}&
   \begin{itemize}[noitemsep,leftmargin=*]
 \item Feature-less environments 
 \item Live updates of changing environment 
 \end{itemize}
 \\ \hline
\textbf{Spatio-temporal Overview}&
  Provide a static overview of movement and gaze trajectories &
  \begin{itemize}[noitemsep,leftmargin=*]
 \item Visualization of gaze and movement trajectories  
 \item Comparison between multiple recordings
 \end{itemize}
 &
  Visual clutter with current approaches \\ \hline
\begin{tabular}[c]{@{}l@{}}\textbf{Dynamic}\\\textbf{Surroundings}\end{tabular}&
  Support for dynamic environments increases range of possible applications and usage scenarios
  &
   \begin{itemize}[noitemsep,leftmargin=*]
  \item Tracking the environment 
  \item Including trajectories of AOIs in the analysis
  \end{itemize}&
   \begin{itemize}[noitemsep,leftmargin=*]
  \item External sensors necessary 
  \item Limited scalability for environmental scale
  \end{itemize}
  \\ \hline

    \textbf{Environmental Scale}
 &
  Availability of SV across a wide range of environments, e.g., from a small room to a large city &
  All previous challenges become more difficult with a larger environmental scale &
   \begin{itemize}[noitemsep,leftmargin=*]
  \item Handling of large live data problematic 
  \item Uncontrolled conditions impair comparison between participants
  \end{itemize}
  \\ 
\end{tabular}%
}
\end{table*}

General descriptive statistics, for instance, fixation count and duration~\cite{Holmqvist2011} can provide information about differences in viewing behavior. These statistics typically lack the interpretation of when and where people looked at a stimulus in detail.
Other established methods often abstract the stimulus into areas of interest (AOIs), providing semantic meaning, but often neglecting the spatial dimension~\cite{Blascheck2017} by reducing fixations on AOIs to labels.

For situated visualizations, the spatial context is the main aspect that differs from typical desktop applications. Hence, the context should be preserved in the analysis. The combination of position and gaze tracking then allows investigating questions regarding spatial aspects, such as: 
\begin{center}
    ``\textit{Where are people standing when they look at specific parts of a visualization?}'' 
\end{center}

A question that is very important for scenarios in high-risk areas, for example, in manufacturing halls (see Section~\ref{sec:Scenarios}).
Overall, the advantages of eye tracking for situated visualization comprise aspects also apparent in desktop scenarios
and include spatial aspects that become important in respective evaluation scenarios.

\section{Challenges for Eye Tracking in AR}\label{sec:Challenges}
\begin{figure*}[h!t]
    \centering
    \begin{subfigure}[b]{0.49\textwidth}
    \centering
        \includegraphics[height=0.48\linewidth]{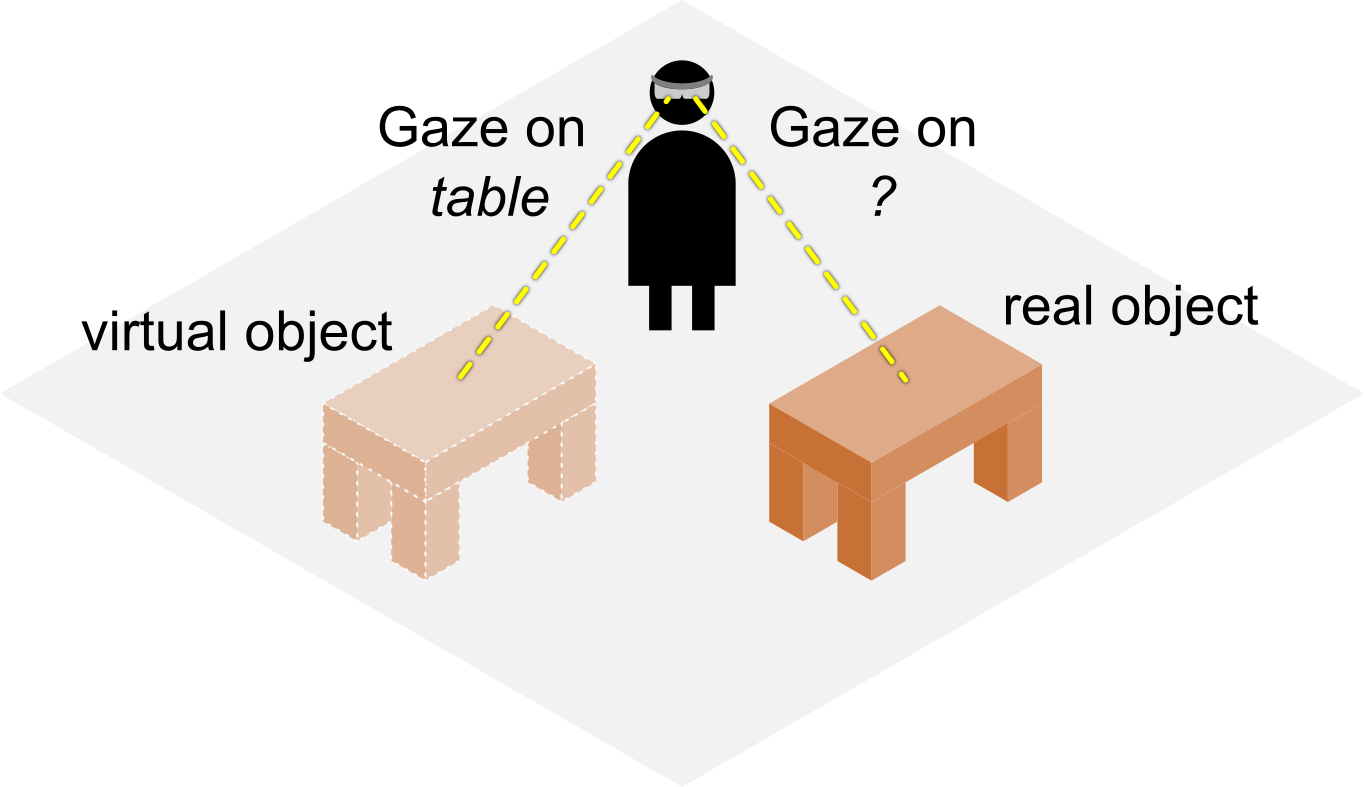}
    \caption{Semantic mapping of virtual content can be derived automatically, real objects pose a classification problem.}
    \label{fig:challenge1}
    \end{subfigure}
    \hfill
    \begin{subfigure}[b]{0.49\textwidth}
    \centering
        \includegraphics[height=0.48\linewidth]{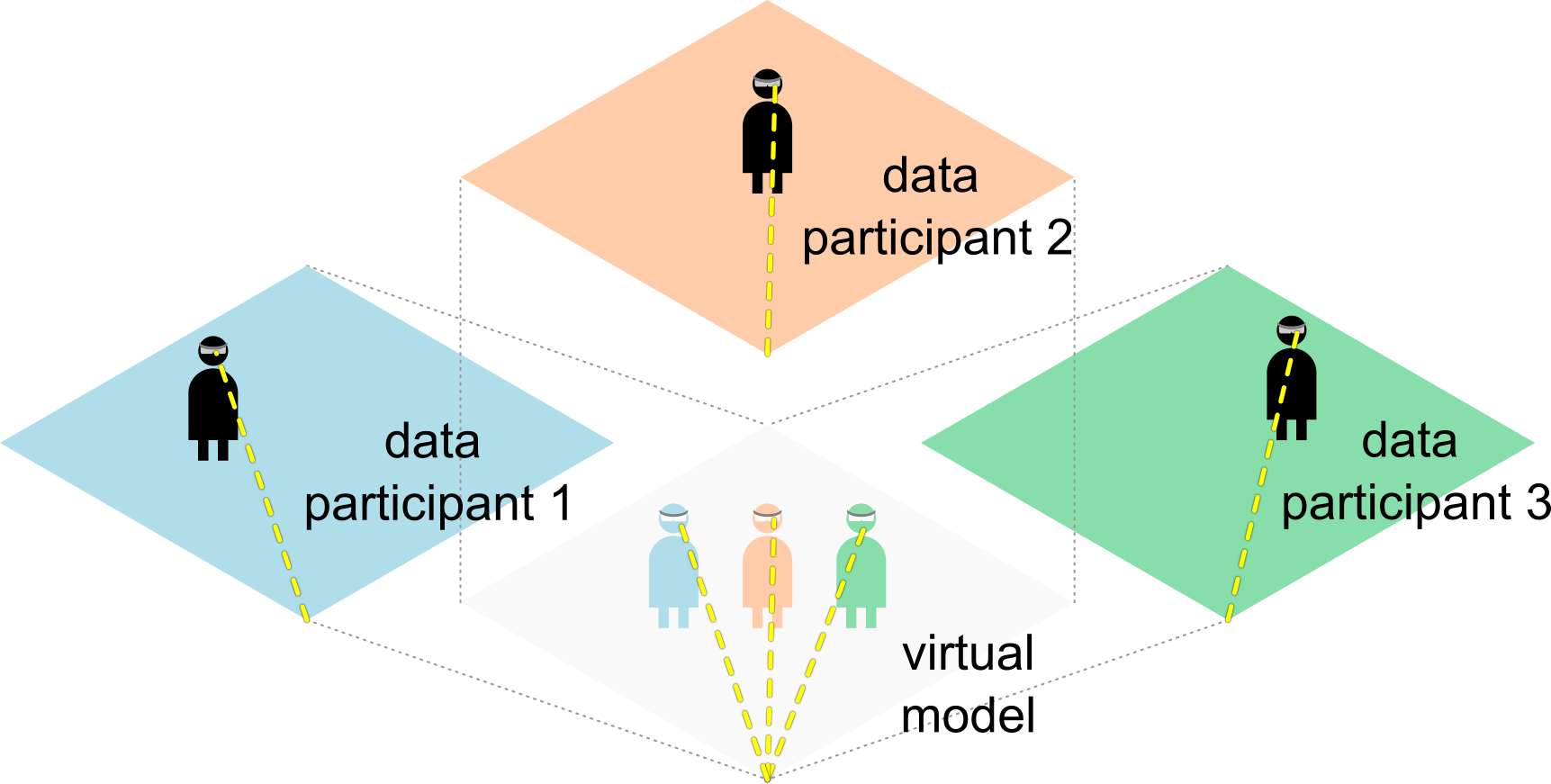}
    \caption{Spatial mapping of individual recordings and additional data (e.g., 
 3D models) into a common coordinate system.}
    \label{fig:challenge2}
    \end{subfigure}
    \vspace{2ex}
    
    \begin{subfigure}[b]{0.49\textwidth}
    \centering
        \includegraphics[height=0.48\linewidth]{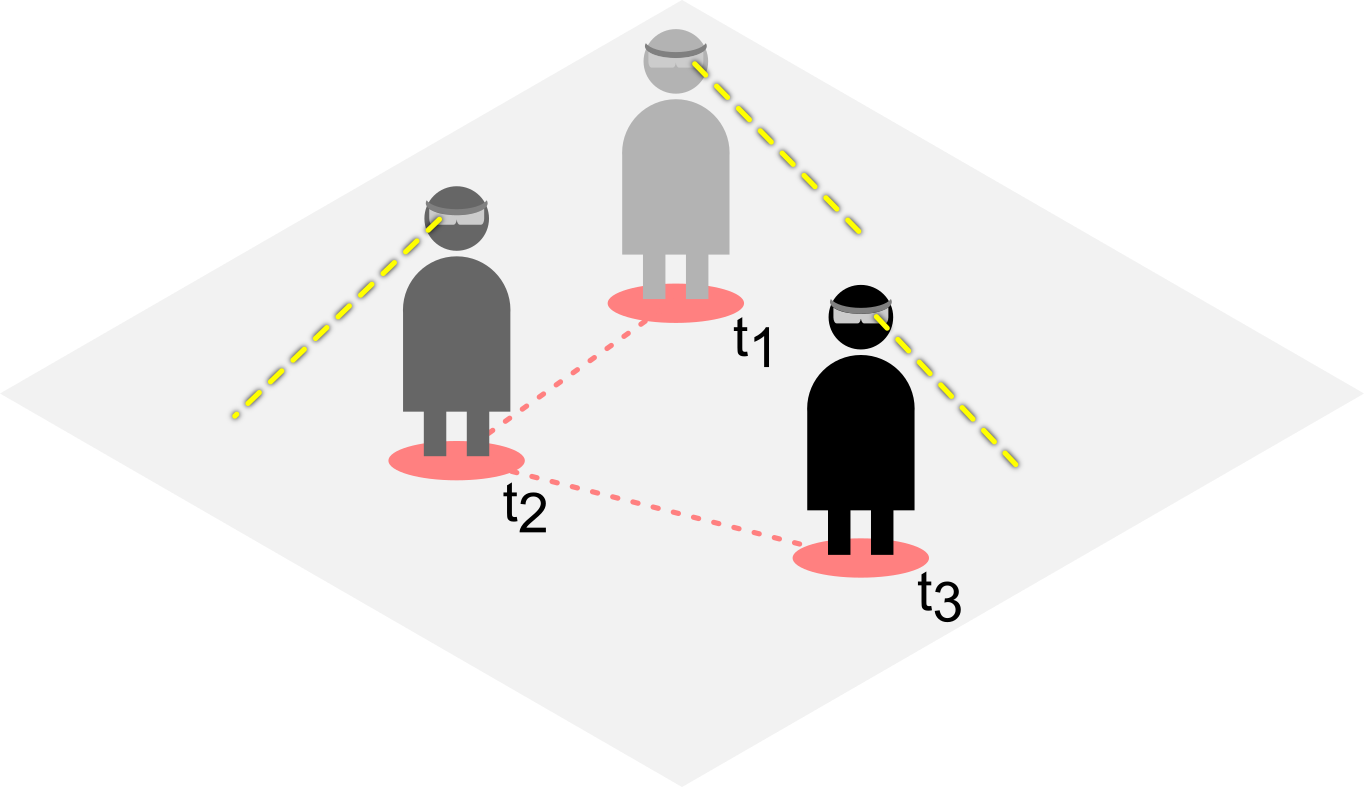}
    \caption{Spatio-temporal overview of the recorded movement and gaze trajectories of the user.}
    \label{fig:challenge3}
    \end{subfigure}
    \hfill
    \begin{subfigure}[b]{0.49\textwidth}
    \centering
        \includegraphics[height=0.48\linewidth]{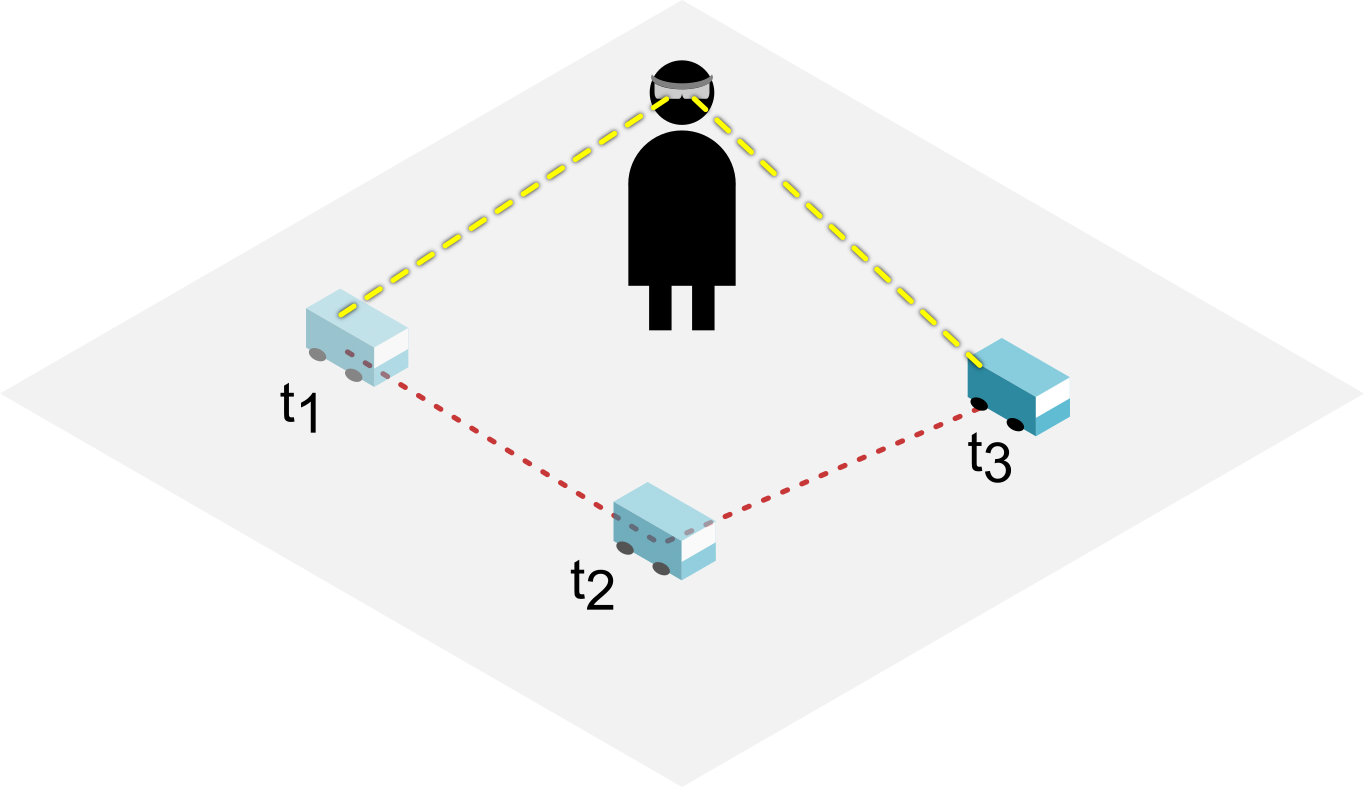}
    \caption{Dynamic surrounding objects also follow spatio-temporal trajectories.\\}
    \label{fig:challenge4}
    \end{subfigure}
    \vspace{2ex}
    
    \begin{subfigure}[b]{0.99\textwidth}
    \centering
        \includegraphics[height=0.25\linewidth]{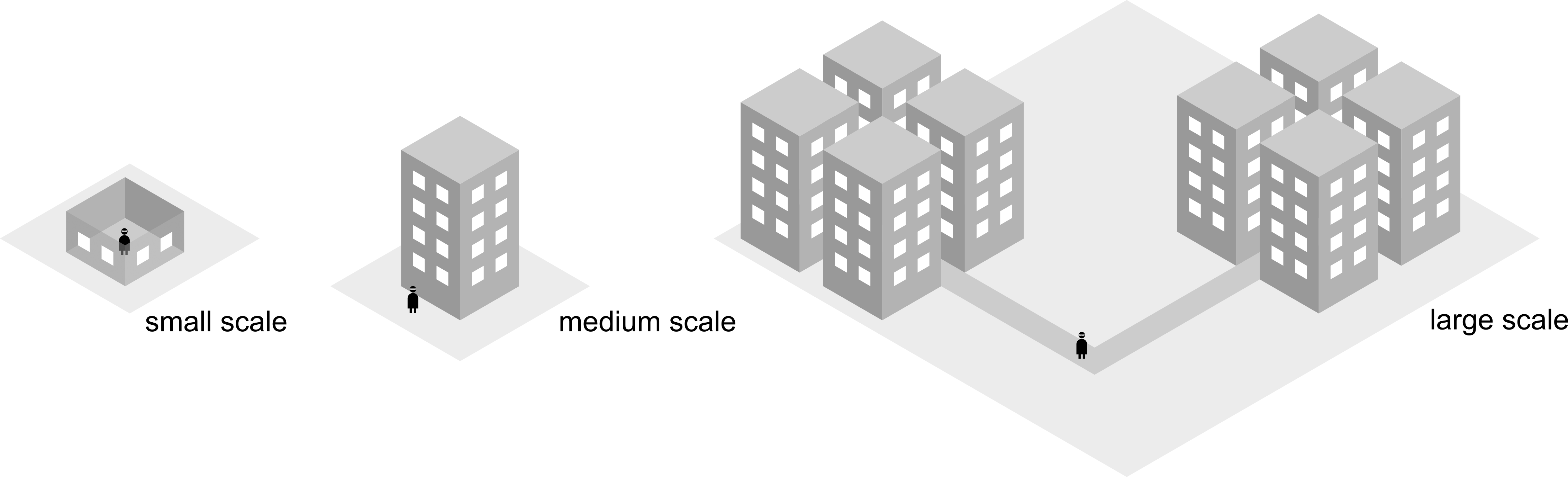}
    \caption{Environmental scale increases the difficulty of all previous challenges respectively. Small-scale experiments in single rooms or floors will be sufficient for many applications, but different scenarios could also benefit from medium scale (e.g., multi-story buildings) up to large-scale environments (e.g., cities).}
    \label{fig:challenge5}
    \end{subfigure}
    
    \caption{Challenges for eye tracking in augmented reality comprise semantic mapping of gaze, spatial mapping of multiple participants, spatio-temporal overview of egocentric and surrounding recordings of movement, and the environmental scale of the scenario.}
    \label{fig:challenges}
\end{figure*}

With AR, new challenges for the acquisition, processing, and representation of gaze data arise (Table~\ref{tab:challenges_summary}). We mainly have to differentiate between virtual and real content, how different recordings are matched, and address multiple challenges concerning the analysis of the recorded spatio-temporal data (Figure~\ref{fig:challenges}).

\subsection{Semantic Mapping}

One common problem with eye tracking analysis is the interpretation of a sequence of consecutive fixations, mainly focusing on questions \textit{what} was investigated. The definition of AOIs is often necessary to provide a semantic mapping of gaze to specific objects or behavior patterns. If this information is available, scanpaths and comparisons between participants become better interpretable and can be visualized in multiple ways~\cite{Blascheck2017}.
To derive this semantic information, we have to differentiate between virtual and real-world content (Figure~\ref{fig:challenge1}).

Virtual content is rendered under the control of the developers/study designers. Hence, the semantic interpretation of virtual objects can be derived automatically if necessary information such as labels is provided in advance. This aspect should be considered during the design of the virtual content and included in visualization pipelines for augmented reality~\cite{Zollmann2020}.

For all real-world content, similar issues arise as with eye tracking with mobile glasses, i.e., world knowledge becomes essential for data analysis. 
If AOIs are defined on a per-object basis, pre-trained classifiers~\cite{Wolf2018} and unsupervised clustering~\cite{Barz2020} help interpret and label gaze on AOIs. In the worst case, manual annotation~\cite{Kurzhals2021} is necessary for individual fixations or gaze samples to assign appropriate labels, which is one of the most time-consuming steps in eye tracking analysis. We expect that future approaches to address this issue will be based on video input from the HMD.
Further, semantic segmentation methods for the underlying 3D mesh~\cite{Valentin2013} could also provide important information about static AOIs in the environment.

\subsection{Spatial Mapping}
Evaluation procedures aim for a description of general behavior patterns. Hence, the continuously recorded position of people and their gaze have to be mapped into a common 3D coordinate system for comparative analyses. Currently, devices such as the HoloLens2 provide this capability based on environmental features. Additional components, such as meshes from photogrammetry or laser scans can be integrated to provide more details of the spatial context, but also require a correct embedding in the common 3D space.

Simultaneous Localization and Mapping (SLAM)~\cite{Cadena2016} in AR devices allows estimating the position of a user in an unknown environment and relating it to a virtual model, which is generated in real-time. AR devices equipped with an Inertial Measurement Unit (IMU) can make the tracking of users even more robust \cite{Feigl2020} and allow tracking the relative displacement of users to a referenced pose \cite{Zendjebil2008}. 
However, the SLAM algorithm does not provide a reference to a global position, such as GPS positions do~\cite{Reitmayr2010}. Therefore, the recording of motion and gaze data from different users will generate data with different reference coordinate systems (see  Figure~\ref{fig:challenge2}). In order to compare the data of different users, a common coordinate system is needed. This can be achieved by defining an absolute coordinate system, to which the relative coordinate systems are aligned to. Even though SLAM provides robust localization, it comes with certain limitations. 
Problems include feature dependency, scaling and mapping errors, and inaccuracies at dynamic object motion~\cite{Feigl2020}. Consequently, this means that an alignment of multiple models is possible as long as sufficient features are available.

Spatial mapping allows the alignment of virtual objects onto real-world objects (walls, tables, etc.). However, the limitations introduced in spatial mapping and the update of the spatial map in each frame may cause the virtual objects to drift away from their original position \cite{Ong2017}. In order to stabilize the virtual objects in the real world, spatial anchors can be utilized.
One common way to achieve this requires users to place spatial anchors manually in the scene, which cause a persistent placement of virtual objects over time and multiple AR sessions \cite{Bachras2019}. 

For small-scale scenarios (Section~\ref{subsec:scale}), current solutions provide sufficient mapping in feature-rich environments
to compare recorded trajectories of multiple people in one common coordinate system. Spatial mapping in large-scale outdoor scenarios proves more difficult and is related to the challenges researchers on autonomous driving face.

\subsection{Spatio-temporal Overview}
Continuous recording of position and gaze direction results in two trajectories of the respective data (Figure~\ref{fig:challenge3}). 
Gaze direction is partially constrained by the position (as a function of the real world), but position is usually continuous, whereas fixations are not.
This complex type of data, i.e., spatio-temporal data in 3D space, is challenging to visualize in an overview. 

Current approaches display a 3D model of the recorded scene and represent movement trajectories as lines and gaze attached with directional clues~\cite{Han2021}, or as heat map on reconstructed surfaces~\cite{Jogeshwar2021}, similar to Figure~\ref{fig:teaser}. Animated replays of the data are also common, allowing to re-investigate the path of a participant and what the respective person looked at from an ego-centric point of view, or an arbitrary point in the scene.
However, these approaches do not scale well with an increasing number of participants and longer recordings where different positions will be visited multiple times. They will, similar to trajectory visualizations or vector field visualizations in 2D or 3D, lead to visual clutter due to overdraw and line crossings. 

We see a need for new visualization techniques to support this type of analysis. AOI-based techniques~\cite{Blascheck2017}, for example, scarf plots which abstract scanpaths to color-coded timelines, could be re-combined with spatial representations to provide a comprehensive framework for the analysis of numerous participants. 

\subsection{Dynamic Surroundings}
In addition to the dynamic movement of the person wearing the HMD, the environment can also change (Figure~\ref{fig:challenge4}). For instance, other people, moving objects or changing environmental conditions that are not part of the virtual world model
have to be addressed separately if required by the analysis task. 

Mapping gaze to dynamic objects can be challenging. They occur at different locations and different points in time which is hard to capture with static models of the environment as mentioned before. Oishi et al.~\cite{Oishi2021} address this issue by presenting a framework called \textit{4D Attention} which maps gaze to static and dynamic objects. While for static objects a pre-built 3D model of the environment is utilized for gaze mapping, the dynamic objects are reconstructed on the fly and gaze is mapped accordingly. Their approach differentiates between rigid and non-rigid dynamic objects for reconstruction.
Since rigid objects have a fixed posture without any deformation, their 6-DoF pose can be determined and tracked in real time. The non-rigid objects, however, change their posture and go through a deformation process requiring different measures for 3D reconstruction.   
While this approach provides valuable information during replay and in real time, it also focuses on single participants.
By including information about dynamic changes in the environment, the difficulty for the analysis increases significantly, as such events are much harder to track than the person wearing the HMD. Similar to the issues discussed for semantic mapping, we expect that parts of this problem can be addressed in the future by the visual input from the optical devices in the HMD, for instance, object identification and pose estimation which will then be integrated into the trajectory data.
Additionally, external tracking devices (e.g., GPS) could also support the mapping of gaze to dynamic objects and record their respective trajectories.

\subsection{Environmental Scale} \label{subsec:scale}
Depending on the scenario, the scale of the environment may vary significantly. In confined spaces, e.g., in a gallery, indoor navigation provided by the HMD in combination with spatial mapping is possible with current generation devices. Figure~\ref{fig:teaser} shows an example of a single room that was reconstructed offline 
as a reference model.
We refer to this as a \emph{small scale scenario} (Figure~\ref{fig:challenge5}). An extension to multiple rooms and stories would complicate representation and tracking and could be considered a medium-scale scenario.

Outdoor AR finds application in the military domain, civil engineering, cultural heritage, entertainment, and more~\cite{Zendjebil2008}. The augmentation of urban environments with virtual content can support a user's understanding of an environment.  The embedding of visual cues into the surroundings to support navigation and find target locations \cite{Hoellerer1999} is also a common outdoor AR application. Including eye tracking in such environments can be used to improve UI design and analyze the perception of virtual content and its influence on spatial cognition. For example, in navigation-based AR applications \cite{Josephson2019}, or to study the attention of users in the presence of visual distractors while driving to improve safety \cite{Chatterjee2020}. 

Extending measures to scenarios with outdoor navigation is especially challenging. It will require additional localization techniques, potentially derived from fusion with other sensors (e.g., GPS).
The complexity of the world model in such a \emph{large scale scenario} can potentially invalidate approaches from the smaller scale both because of data handling and algorithmic scalability.
Work on localization algorithms on this scale~\cite{Schoenberger2018} could also help, but detailed models of all investigated content will be hard to obtain. Accordingly, capturing the dynamic surrounding becomes a goal that seems hard to achieve with current approaches. 

\begin{figure*}[t]
    \centering
    \includegraphics[width=\textwidth]{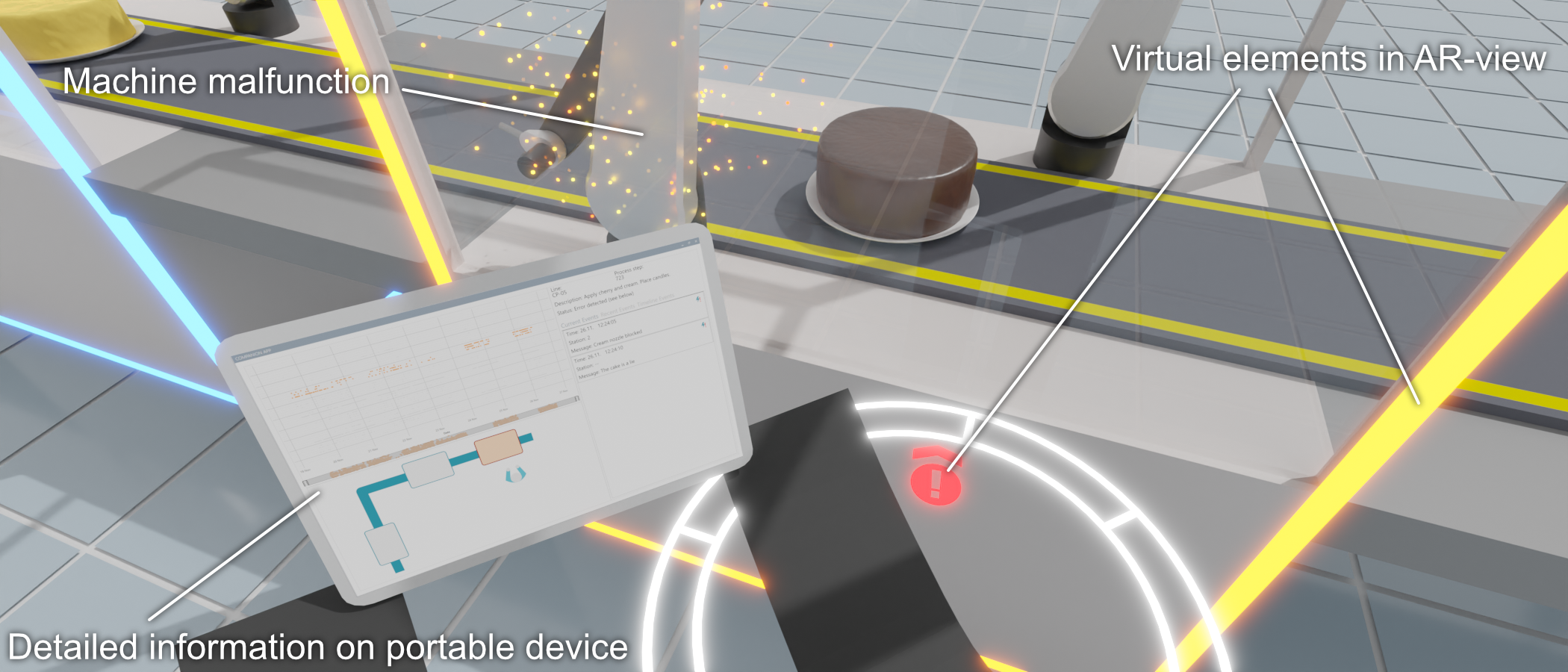}
    \caption{Illustration of an existing setup for situated monitoring and analysis in a manufacturing environment (see Becher et al.~\cite{Becher2022} for images of the running real-world prototype). The method combines AR situated visualizations, including a virtual compass that shows ongoing error events and visual highlighting of machine tools. A portable tablet computer displays more detailed views for visual analytics. The devices share data in real-time, such as the current position of the operator on the shop floor.}
    \label{fig:manufacturing}
\end{figure*}

\section{Usage Scenarios}\label{sec:Scenarios}
The presented usage scenarios have been selected either because situated visualizations have been applied previously, or because we see much potential for their application.
This mainly comprises scenarios with live machine and sensor data processing in manufacturing and architecture and more informative visualizations in the context of education and personal life-logging.

\subsection{Manufacturing in Industry 4.0}
In modern Industry 4.0 environments, manufacturing has become increasingly automated and sensor-equipped machine tools produce a wealth of data.
Operators on the shop floor perform fewer manual tasks and instead need to monitor and analyze machine data in order to quickly take action in case of malfunctions.
Becher~et~al.~\cite{Becher2022} presented a situated analytics approach to support operators in real-time.
They propose to use a combination of situated visualizations presented via HMD and visual analysis tools available on a portable tablet computer (see Figure~\ref{fig:manufacturing}).
Evaluation was limited to expert interviews using the System Usability Scale (SUS) and questionnaires.

A central concept of situated visualizations is context sensitivity.
This includes limiting displayed information to only what is relevant in the current spatial context. Here, eye tracking would allow for more precise filtering that is not only dependent on a user's location but also on the areas that are currently focused.
If historical gaze data from previous faults is available, it is possible to overlay heat maps on the real environment that highlight potentially critical areas on machines that were examined in previous inspections. 

For retrospective analysis of the application, gaze data yields additional insights into application usage.
For example, it is possible to detect when the tablet application is investigated, specifically when it is used to either read displayed information or check the map.
Similarly, it is possible to find out if operators still need to look up additional information from the traditional displays available at the factory. Such displays are often located at machines or status displays installed on the factory ceiling.
Lastly, analysis of the gaze data could reveal insights into visual clutter issues or potential distractions that can be caused by using extensive AR overlays.

\subsection{Architectural Design and Construction}
The application of AR for processes in architecture from the design to the maintenance of a building became feasible in recent years~\cite{Noghabaei2020}. 
With increasing sensor input and consistently updated model data from Building Information Modeling (BIM), situated visualization has the potential to facilitate individual procedures in this context significantly and can map the respective components to their real-world counterparts.
Typical examples include the extension of construction sites through virtual content, depicting the previous state of the site \cite{Gleue2001}, or hidden objects like pipes in the underground\cite{Mendez2006}. 

A comprehensive analysis of a multitude of sensors would benefit from a situated approach.
Considering the challenges (Section~\ref{sec:Challenges}), dynamic changes in the surroundings mainly concern construction processes, similar to the previous example.
However, buildings that are adaptive can further complicate understanding of the current state of the real world, for example when features of a facade change to compensate for wind or solar irradiance~\cite{voigt2022characteristics}.
Given the necessarily high grade of automation of such a building, the respective BIM can offset these challenges. We would expect it to expose these dynamic features as well as provide high-quality data regarding the static components that obviate a great part of the spatial understanding challenges required in traditional settings.

We see the main difference to the manufacturing scenario in the scale of the surrounding context which potentially increases with multiple building complexes and on a vertical scale with multi-story buildings.
Going forward, the consistent and unconditional implementation of BIM can potentially make situated analytics scenarios for architecture comparatively easy to support, since the available annotated CAD data will provide a comprehensive basis for the interpretation of fixations.
Developing SV for a large-scale scenario including multiple buildings and their respective spatial context is still quite a challenge for future research and will raise the question of how to evaluate such an approach. We see a combined tracking of position and gaze as an important supplement to established methods to evaluate how people use situated approaches, identify potential flaws, and provide guidance to avoid safety hazards.

\subsection{Educational Visualization}
Expanding the scope of education methods through augmented reality helps engage the audience in a topic~\cite{Sailer2019}. For example, the addition of virtual content to an exhibited object in a museum can provide additional details, making it more attractive to study. By analyzing eye movements of the visitors, their perceptual and cognitive processes while investigating such hybrid environments could be studied. 
Such analyses could also help improve presenting and conveying information to an audience. 

The behavior of persons in augmented environments, including a museum, was examined by Muchen and Tamke~\cite{Muchen2021}. The attention of participants was approximated by head poses and visualized by obtaining a voxel representation of the environment and coloring the voxels based on the intensity of attention received, leading to a heat map visualization. Further, the areas of interest and fixation sequence were displayed with the voxels.
The evaluation of the behavior data collected in the museum revealed the visiting time for different art pieces and also the preferences of individuals. 
The movement data indicated how the space was examined and in which order the different locations were visited. 
Based on the visualization of this information, the aim of the work was to
\textit{``...provide insights to design problems that are overlooked in retrospective evaluations, and therefore architects can better engage users who are not professionally trained in the design phase...''}~\cite{Muchen2021}. 
To improve the understanding of the behavior patterns, the visualization could be extended with a gaze replay, which helps analyze the data of the individuals at different timespans and therefore can provide more details about decisions made at each point in time.

Another use case introduced by Lu et al.~\cite{Lu2020} applies eye tracking to detect problems during surgical training. The authors developed a system that displays instructions on an AR display whenever the user faces difficulties during the simulation of the surgery.

Based on the given examples, we see many educational scenarios as small to medium-scale environments where virtual and real-world content can be clearly defined. Changes in the surrounding could be part of the scenario but are less problematic than in uncontrolled outdoor scenes.

\subsection{Personal Visualization}
As a more futuristic scenario, HMDs could 
also be applied as a life-logging device to create personal visualization~\cite{Huang2014} providing information about personal activities, health functions, etc. The integration of eye tracking for personal data was also suggested in the past for personal encounter recaps~\cite{Kurzhals2015}. AR technology could also provide immediate feedback on attention-related aspects. For example, based on multi-user data, attention could be guided to both popular and hard-to-identify sights on a hiking trip where tourist signs are less prominent (or less frequent) than in cities.

This scenario raises numerous questions concerning data privacy~\cite{Goebel2020}, especially if not only the personal data would be recorded, but also the environment. One aspect is that people record their own data including locations and attention. Potential misuse is possible if the data is handed to external parties. If the surrounding is also included, other people might be involved without their consent. Hence, anonymization and data reduction will play an essential role in such scenarios.
Overall, this example showcases that besides technical obstacles, some scenarios will also require special sensibilization with respect to data handling.

\section{Conclusion}
We discussed the advantages and challenges of eye tracking as a means for the evaluation of situated visualization. 
Recording gaze distributions as an indicator for visual attention and fixation sequences for the interpretation of visual strategies have proven to be valuable for the evaluation of visualizations. Established eye tracking metrics and scanpath analysis methods can partially be adapted for augmented reality applications. However, this scenario requires extending the analysis of the spatial dimension of the visual stimulus into 3D, which provides interesting new ways to look at visualization with respect to spatial aspects:
\begin{itemize}
    \item \textit{How do people move and look at situated visualizations?}
    \item \textit{What is the spatial coverage of positions and movement when people use a visualization?}
    \item \textit{What are optimal positions to investigate a visualization?}
\end{itemize}

Data acquisition is already possible with current-generation hardware. However, the analysis of situated visualization scenarios will require further research on how to combine and represent the tracked data from multiple participants best to derive meaningful insights. The addition of eye tracking in future evaluation procedures for situated visualizations is a valuable supplement to established methods of performance measures and qualitative evaluation by interviews and questionnaires.

Regarding the limitations, the challenges mentioned in Section~\ref{sec:Challenges} require an interdisciplinary research effort on how to capture and process the necessary data, as well as new methods to analyze and make sense of the resulting gaze distributions and sequences. However, some inherent limitations of eye tracking will remain: (1) There is no point-precise measurement of the point of regard, uncertainty is always present in the range of the accuracy and precision of the eye tracker. Hence, small parts of a visualization (e.g., single dots, lines) are hard to identify individually. This affects both display-based visualizations as well as small-scale physical models used for data physicalization. (2) Eye tracking provides an approximation for visual attention; however, long recordings will most likely contain periods of mind wandering and stares without attentional focus that will be hard to identify solely based on gaze measures.

Overall, we see the additional collecting of gaze data in the context of a user study as worth the effort to gain a potentially information-rich source to derive insights.
To harness the full potential of eye tracking for evaluating situated visualization and AR applications in general, new techniques will be necessary to process and represent the recorded data appropriately for quantitative and qualitative analyses.

\acknowledgments{
This work is supported by the Deutsche Forschungsgemeinschaft (DFG, German Research Foundation) under Germany's Excellence Strategy -- EXC 2120/1 -- 390831618 and SFB~1244 – Project ID 279064222.
}

\bibliographystyle{abbrv-doi}

\bibliography{template}
\end{document}